\documentclass[draft,nofootinbib,aps,prl,twocolumn,showpacs,groupaddress,preprintnumbers,floatfix]{revtex4-1}

\usepackage{dynlearn}
\usepackage{braket}
\usepackage{multirow}
\usepackage{cleveref}
\newcommand{\ketbra}[1]{\ket{#1}\bra{#1}}

\newcommand{\Hmax}[2]{\mathrm{H}^{#1}_{\max}\left[#2\right]}
\newcommand{\Hmin}[2]{\mathrm{H}^{#1}_{\min}\left[#2\right]}
\newcommand{\Hq}[1]{\mathrm{H}_{q}\left[#1\right]}
\definecolor{mygray}{gray}{0.6}

\theoremstyle{plain}    

\begin{document}

\def\ourTitle{
Thermal Efficiency of Quantum Memory Compression}

\def\ourAbstract{
Quantum coherence allows for reduced-memory simulators of classical processes.
Using recent results in single-shot quantum thermodynamics, we derive a minimal
work cost rate for quantum simulators that is quasistatically attainable in the
limit of asymptotically-infinite parallel simulation. Comparing this cost with
the classical regime reveals that quantizing classical simulators not only
results in memory compression but also in reduced dissipation. We explore this
advantage across a suite of representative examples.
}

\def\ourKeywords{
stochastic process, hidden Markov model, \texorpdfstring{\eM}{epsilon-machine}, causal states, mutual information.
}

\hypersetup{
  pdfauthor={James P. Crutchfield},
  pdftitle={\ourTitle},
  pdfsubject={\ourAbstract},
  pdfkeywords={\ourKeywords},
  pdfproducer={},
  pdfcreator={}
}

\author{Samuel P. Loomis}
\email{sloomis@ucdavis.edu}

\author{James P. Crutchfield}
\email{chaos@ucdavis.edu}
\affiliation{Complexity Sciences Center and Physics Department,
University of California at Davis, One Shields Avenue, Davis, CA 95616}

\date{\today}
\bibliographystyle{unsrt}

\title{\ourTitle}

\begin{abstract}
\ourAbstract
\end{abstract}



\preprint{\arxiv{1911.00998}}

\title{\ourTitle}
\date{\today}
\maketitle

\setstretch{1.1}

\listoffixmes

This Letter demonstrates the potential for a quantum machine to perform a
particular task---namely, simulating hidden Markov models---using not only less
memory storage, but also reduced thermodynamic dissipation. To do so it
synthesizes recent developments in quantum memory compression
\cite{Gu12a,Maho16a,Pals17a,Joun17a,Bind17a,Thom17a,Riec15b,Thom18a,Loom18a,Liu19a}
and quantum thermodynamics
\cite{Janz00a,Vinj16a,Lost18a,Horo13a,Bran13a,Bran13b,Lost15a,Alha16a,Dahl11a,Rio11a,Fais15a,Gury20a}
into a fruitful new crossover framework for studying the thermodynamics of
quantum simulators for stochastic processes.

We begin by reviewing \emph{computational mechanics}, which seeks to analyze
how natural systems manipulate information to produce and transform stochastic
processes \cite{Crut88a,Crut08a,Jame10a,Trav12a,Crut12a}. More recently,
computational mechanics examined the thermodynamics of computation,
generalizing Landauer's principle for memory erasure \cite{Land61a} to derive
the \emph{information processing Second Law} \cite{Boyd15a}, which gives the
minimal cost of transforming one stochastic process in to another, and the
\emph{thermodynamics of modularity} \cite{Boyd17a}, which determines the
implementation costs for transformations.  An important result in this
classical regime is that the Shannon-entropy Landauer bound on average work
cost for a computation can be achieved for a single implementation of that
computation \cite{Boyd17a}. That is, the (quasistatic) single-shot cost of a
computation is the same as the cost for asymptotically-infinite parallel
implementations.

We next consider the relationship between memory and thermodynamics in the
quantum regime. Quantum computational mechanics recently explored how to
simulate and transform \emph{classical} stochastic processes using
\emph{quantum} systems \cite{Gu12a,Maho16a,Thom17a,Bind17a}, even constructing
experimental implementations \cite{Pals17a,Joun17a}. Generally, quantum
simulators of complex processes require less memory (measured by the
quantum-state von Neumann entropy) than classical (measured by the statistical
complexity---classical-state Shannon entropy) \cite{Riec15b,Thom18a}.

Quantum thermodynamics \cite{Vinj16a}, though recently advancing via thermal
resource theories \cite{Janz00a,Horo13a,Bran13a,Lost15a,Lost18a} and
single-shot thermodynamics \cite{Dahl11a,Rio11a,Bran13b,Fais15a}, has not yet
been applied to examine quantum simulators. However, it is known that
Landauer's lower bound, as given in the form of Shannon and von Neumann
entropies, is not generally attainable. A more nuanced view is necessary
\cite{Dahl11a,Bran13b}.  Due to this, single-shot and asymptotic analyses must
be performed separately when transitioning from classical to quantum regimes.

Using these recent results in quantum thermodynamics, we calculate achievable
and lower bounds on the work cost rate for the quasistatic implementation of
quantum simulators, in both the single-shot and asymptotic regimes. These
bounds reveal a direct relationship between memory compression achieved by a
quantum implementation and the increase in extractable work via the same. We
then elucidate the nature of this trade-off across a suite of examples.

\begin{figure}[t]
\centering
\includegraphics[width=\columnwidth]{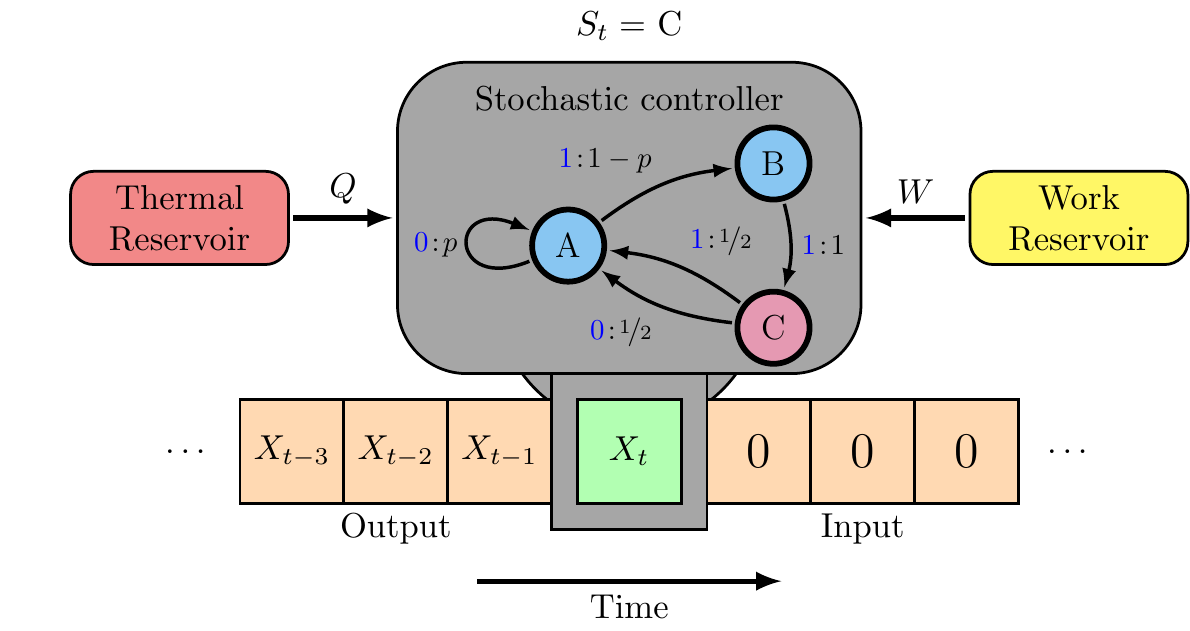}
\caption{Information ratchet sequentially generates a symbol string on an
	empty tape: At time step $t$, $S_t$ is the random variable for the ratchet
	state. The generated symbols in the output process are denoted by
	$X_{t-1},X_{t-2},X_{t-3}, \ldots$. The most recently generated symbol $X_t$
	(green) is determined by the internal dynamics of the ratchet's memory,
	using heat $Q$ from the thermal reservoir as well as work $W$ from the work
	reservoir. \emph{(Ratchet interior.)} The memory dynamics and symbol
	production are governed by the conditional probabilities
	$\Pr(s_{t+1},x_t|s_t)$, where $s_t$ is the current state at time $t$, $x_t$
	is the generated symbol and $s_{t+1}$ is the new state. Diagrammatically,
	this is a hidden Markov model---a labeled, directed graph in which nodes are
	states $s$ and edges represent transitions $s\rightarrow s'$ labeled by the
	emitted symbol and associated probability ${\color{blue} x}:\Pr(s',x|s)$.
  }
\label{fig:Ratchet}
\end{figure}

{\bf Ratchets, generators, and processes.} 
By \emph{computation} we mean transducing an input sequence $\dots x_0 x_1
x_2\dots$ into a new output sequence $\dots y_0 y_1 y_2 \dots$. When a
computation is done \emph{online} the implementation acts on a single input
symbol at a time and immediately determines the output symbol based on
\emph{stored memory} of the past inputs and outputs. The graphical
representation of such a transducer, see Fig. \ref{fig:Ratchet}, suggests
calling them \emph{ratchets}, as done previously \cite{Boyd15a}.

The following restricts itself to \emph{finite generators} that take only a
trivial (constant) input, produce as output a stochastic process taking values
in a finite alphabet $\mathcal{X}$, and employ a finite set $\mathcal{S}$ of
memory states. This is sufficient to explore the central relationships between
memory and thermodynamics.

A finite generator's operation is described by the probabilities
$\Pr\left(s',x\middle|s\right)$ of emitting symbol $x$ and ending in a final
memory state $s'$, if starting in memory state $s$. This is depicted as the
\emph{hidden Markov model} (HMM) \cite{Uppe97a} in the ratchet interior
(Fig. \ref{fig:Ratchet}).

An HMM is \emph{irreducible} if the matrix $\Pr\left(s'|s\right):=\sum_x
\Pr\left(s',x\middle|s\right)$ is irreducible. Functionally, this means that
every state $s'$ can be reached from any other state $s$ in the state
transition diagram.

An irreducible HMM has a unique \emph{stationary distribution} $\Pr_0(s)$ over
its memory states such that $\sum_{s} \Pr\left(s'\middle|s\right) \Pr_0(s)=
\Pr_0(s')$. We let $S'XS\sim\Pr\left(s',x\middle|s\right) \Pr_0(s)$ represent
the joint random variable of the generated symbol and the memory state before
and after generation. Over many time steps, we accrue the joint random variable
$S_{t+1} X_t \dots S_2 X_1 S_1 \sim \Pr(s_{t+1},x_t|s_t)\dots
\Pr(s_{2},x_1|s_1) \Pr_0({s_1})$. The sequence of random variables $X_t \dots
X_1$, typically not independent of each other, describes $t$ samples of the
\emph{stochastic process} simulated by the generator.

Most results on generators and their quantum counterparts encompass those
with at least the two following properties:
\begin{enumerate}
      \setlength{\topsep}{0pt}
      \setlength{\itemsep}{0pt}
      \setlength{\parsep}{0pt}
\item \emph{Predictivity}: $s' = f(x,s)$. The next state is always determined
	by the previous state and the generated symbol: $\Pr(s',x|s) \propto
	\delta_{s',f(x,s)}$ for some $f$. (Elsewhere known as \emph{unifilarity}
	\cite{Ash65a}.)
\item \emph{Minimality}: For any two states $s$ and $s'$, if
	$\Pr\left(x_t\dots x_1 \middle|s\right)=\Pr\left(x_t\dots x_1
	\middle|s'\right)$ for all $t$, then $s=s'$. This ensures that no two
	states predict the same future distributions. A nonminimal HMM is minimized
	by merging predictively equivalent states.
\end{enumerate}
For any process, there is a unique generator satisfying these two properties,
called the \emph{\eM} \cite{Crut12a,Trav12a}.

Given an \eM, from its stationary distribution we can calculate its
\emph{statistical complexity} (\ie, memory) $C_\mu:=\H{S}$ \cite{Crut88a} and
\emph{non-Markovity} $N_\mu := \H{S'|X}$, using the Shannon entropy and
conditional entropy, respectively \cite{Cove06a}.

{\bf Physical implementations.} 
We define a generator's \emph{implementation} as the sextuplet
$\left(\mathcal{H}_S,\mathcal{H}_X,\mathcal{H}_A,\mathcal{H}_B, U,
\mathcal{E}\right)$ consisting of four Hilbert spaces (\emph{memory},
\emph{output}, \emph{auxiliary}, \emph{bath} systems); a unitary $U$ on all
four; and an ensemble $\left\{\ket{\psi_s}:s\in\mathcal{S}\right\}$ embedding
the classical memory states into the memory system $\mathcal{H}_S$,
respectively. The auxiliary system starts in a given pure state $\ket{0}_A$,
while the bath is taken to start in a thermal state. Following convention for
information reservoirs, we consider the memory, output, and auxiliary systems
to be energyless, though the bath system may have some nontrivial Hamiltonian
$H_B$. Under these conditions, we require $[U,H_B]=0$, following the rules for
microscopic energy conservation \cite{Janz00a}. Furthermore, as long as we
begin and end in an information reservoir, we may assume our operations are
performed via Hamiltonian control, with minimal work costs defined by the
state-averaged changes in energy level over a quasistatic erasure of system $A$
\cite{Rio11a,Vinj16a,Alha16a}.

When the auxiliary and the bath are traced out, the implementation must take
the form of the positive map:
\begin{align*}
\mathcal{T} \! \left(\ketbra{\psi_s}\otimes \ketbra{0}\right)  \! = 
 \! \!  \sum_{x,s'}\Pr\left(s',x|s\right) \ketbra{\psi_{s'}} \otimes \ketbra{x}
  .
\end{align*}
Resetting the thermal bath has no associated cost---it may simply be brought
into contact with a larger bath. However, if an auxiliary system is used,
its reset (erasure) cost must be taken into account.

\newcommand{\kB}{k_\text{B}}

{\bf Classical implementation.} 
Two concrete types of implementation have been considered previously. The first
addresses efficiently implementing the generator via classical thermodynamics.
Using Hamiltonian control, Ref. \cite{Boyd17a} showed that \emph{any}
stochastic channel can be implemented in a way that achieves the Landauer
bound. In particular, applying a channel $\Pr(y|x)$ to a random variable $X$,
resulting in $Y$, can be performed with the work cost $W =
\kB T\left(\H{X}-\H{Y}\right)$. For generators this means we can achieve the
work cost per time step of:
\begin{align}
W_\mu = \kB T\ln 2\left( \H{S} -  \H{S'X} \right)
  ~.
\label{eq:classcost}
\end{align}
Written differently:
\begin{equation}\label{eq:classcost-exp}
  W_\mu = \kB T\ln 2\left(C_\mu - N_\mu - \H{X}\right)
  ~.
\end{equation}
This is, in a sense, maximally efficient: The work can never be made lower than
Landauer's bound, so this is the best we can possibly do over all classical
implementations. Noting that $\H{S'}=\H{S}$, we find
$W_\mu = - \kB T\ln 2 ~\H{X|S'} \leq 0$, so predictive generators may
extract positive work.

Much of the cost $W_\mu$ is due to the local nature of the generator: it does
not have access to previously generated symbols to choose its operations.  The
information processing Second Law (IPSL) \cite{Boyd15a}, when applied to
generators, states that the work cost per symbol for generating a process
\emph{with} access to the previous symbols is bounded by:
\begin{align*}
  W \geq -\kB T\ln 2 ~\hmu
  ~,
\end{align*}
where $\hmu = \lim_{t\rightarrow \infty} \frac{1}{t}\H{X_1 \dots X_t}$ is the
\emph{entropy rate} of the process being generated.  Generally, $W_\mu \geq
-\kB T \ln 2 ~\hmu$ \cite{Boyd17a}. 

Classical generators require memory, quantified in the case of the \eM by the
statistical complexity $C_\mu$. Memory cost can be reduced by embedding the
memory states into a nonorthogonal quantum ensemble. This motivates the use of
quantum implementations of generators. 

{\bf Quantum implementations.}
In a quantum implementation, we apply a unitary operator to $SXA$ alone. This
unitary is divided into two parts: $U_{SXA} = \left(1_S\otimes
U_{XA}\right)\left(U_{SX}\otimes 1_A\right)$. The first operation $U_{SX}$
evolves the memory system and the output system to achieve the necessary
correlation. While the second $U_{XA}$ entangles the output and the auxiliary
to represent the effect of a measurement device. The first takes the form:
\begin{align*}
U_{SX}\ket{\psi_s}\ket{0} = \sum_{x,s'} e^{i\phi_{xs}} \sqrt{\Pr(x|s)} \ket{\psi_{f(s,x)}}\ket{x}
  ~,
\end{align*}
where $\ket{x}$ form an orthogonal computational basis representing the
generated symbols. When the generator is predictive, a unitary performing this
transformation exists for any choice of the arbitrary phases $\phi_{xs}$
\cite{Liu19a}.

For a quantum implementation of an \eM, we can measure its memory cost by the
quantum complexity $C_q:= \Hq{S}$ and its quantum non-Markovity as $N_q :=
\Hq{S'|X}$, where $H_q(S)$ is the von Neumann entropy of the stationary state
$\rho_S$ on system $S$ and $\Hq{S'|X}$ is the conditional entropy of the state
$\rho_{SX}'$ after implementing $U_{SX}$. These states have the form:
\begin{align*}
  \rho_S & = \sum_s \Pr{}_0(s) \ketbra{\psi_s} ~\text{and}\\
  \rho_{SX}' & = \sum_{s',x,s} \Pr(s',x|s)\Pr{}_0(s) \ketbra{\psi_s}\otimes \ketbra{x}
  ~.
\end{align*}
For any \eM quantum implementation, we have $C_q\leq C_\mu$, with strict
equality only when the \eM is retrodictive \cite{Maho16a,Loom18a}. 

In general, the single-shot case---that implements a single copy of a
generator---cannot achieve Landauer's bound. Synthesizing several results in
quantum erasure and information processing \cite{Dahl11a,Rio11a,Fais15a}, the
Supplementary Material \cite{QMCATESuppMat} (SM) derives our first main result:
a single generator can be implemented with a work cost of no more than:
\begin{align}
  \begin{split}
  \frac{W_q^{\epsilon}}{\kB T\ln 2} \leq &\ \Hmax{\epsilon^2/4}{S}-\Hmin{\epsilon^2/64}{S'\middle|X}\\
  & \qquad - \Hmin{\epsilon^2/64}{X}
  {+ O\left(\log\frac{1}{\epsilon}\right)}
  \end{split}
  ~,
\label{main-single-shot}
\end{align}
with a probability of failure less than $\epsilon$. Rather than use the Shannon
entropies of the classical work, this is expressed in the smooth conditional entropies of
quantum information theory \cite{Renn04a,Renn05a,Toma13a,Vita13a,Toma2016} as applied to the states
$\rho_S$, $\rho_{SX}'$, and $\rho_X'=\mathrm{Tr}_{S}\left(\rho_{SX}'\right)$.

Suppose, instead of implementing a single copy of a generator, we implement $N$
generators in parallel, each producing an independent realization of the
desired process. The \emph{asymptotic equipartition property} of smooth
entropies \cite{Toma13a} then shows that the work rate $W_q :=
\lim_{\epsilon\rightarrow 0}\lim_{N\rightarrow \infty} W_q^\epsilon/N$ is given
by:
\begin{align}
   \begin{split}
   \frac{W_q}{\kB T\ln 2} &= \Hq{S}-\Hq{S'\middle|X}
   - \H{X}\\
   & = C_q - N_q - \H{X}
   \end{split}
  ~.
\label{main-asymptotic}
\end{align}
where $\Hq{\cdot}$ is the von Neumann entropy \cite{Niel10a}. 
$C_q$, $N_q$, and $W_q$ are functions of the quantum implementation chosen; in
particular, $W_q = W_q(\phi_{xs})$ is a function of the phases. The SM shows
that this is always at least as small as the classical cost: $W_q(\phi_{xs})
\leq W_\mu$. Combining this with the IPSL and $W_\mu$'s negativity gives:
\begin{align}
   - \kB T\ln 2 ~\hmu \leq W_q(\phi_{xs}) \leq W_\mu \leq 0
  ~,
\label{inequality}
\end{align}
for all $\{\phi_{xs}\}$.
Thus, the quantum implementation of a predictive generator offers improvement
over the classical implementation in the work that can be thermodynamically
extracted.

\Cref{main-single-shot,main-asymptotic,inequality} are our three primary
results. In the remainder, we explore in a suite of example generators the
relationship between the \emph{memory compression} $\Delta_q C:= C_\mu - C_q$
and the \emph{work advantage} $\Delta_q W := W_\mu - W_q$. (The suite covers
the qualitatively distinct behaviors observed in our numerical exploration.) We find that the
\emph{efficiency of compression} $e_q = \Delta_q W/\left(\kB T\Delta_q C \ln
2\right)$---the improvement in work cost for each bit of compression
achieved---is a key quantity for monitoring the behavior of quantum
implementations.

{\bf Markov Generators.}
A Markov chain $X_1 \dots X_t$ is a chain of random variables $X_t$, where each
variable is conditionally independent of the past given its predecessor:
$\Pr(x_t | x_{t-1} \dots x_0) = \Pr(x_t|x_{t-1})$ for all $t$. In a sense, a
Markov chain is its own generator---one in which memory states are also the
produced symbols: $\mathcal{S} = \mathcal{X}$. For Markov generators, which
type has historically dominated physical modeling, knowing the produced symbol
$X_t$ automatically determines the next state $S_{t+1}$, as they are identical.
Their non-Markovity $N_\mu=0$ (hence the name of that quantity) and $C_\mu =
\H{X}$. 

As a consequence, the relationship between memory compression and work
advantage is particularly direct. Classical work extraction is simply $W_\mu =
\Cmu - \H{X} = 0$, indicating that Markov chain generation is thermodynamically
neutral at best. However, $W_q = C_q - \H{X} \leq 0$, such that quantally
compressed Markov chain generators are indeed capable of work extraction.  The
memory and work advantages also take on a simple relationship: $\Delta_q W =
\kB T\Delta_q C$ and so they are maximally efficient: $e_q = 1$.

{\bf $R,k$-Golden Mean Hidden Generators.}
However, measurements are typically not a process' internal state; thus, we
must address hidden Markov generators. In addition to non-Markovity $N$,
another means of quantifying how distant a process' generator is from being
Markov is the \emph{Markov order}: the smallest integer $R$ such that
$\H{S_R\middle|X_{R-1} \dots X_1} = 0$. In other words, $R$ is the largest
number of symbols we must see before the generator's state is known. It is
infinite for most processes \cite{Jame10a}; for Markov-generated processes
$R=1$.

A dual notion to the Markov order is the \emph{cryptic order}. This is the
smallest integer $k$ such that $\lim_{t\rightarrow \infty}\H{S_k\middle|X_1
\dots X_t} = 0$. This is a more general condition than that for Markov order:
consequently, $k\leq R$ for all processes.

\begin{figure}[t]
\centering
\includegraphics[width=0.9\linewidth]{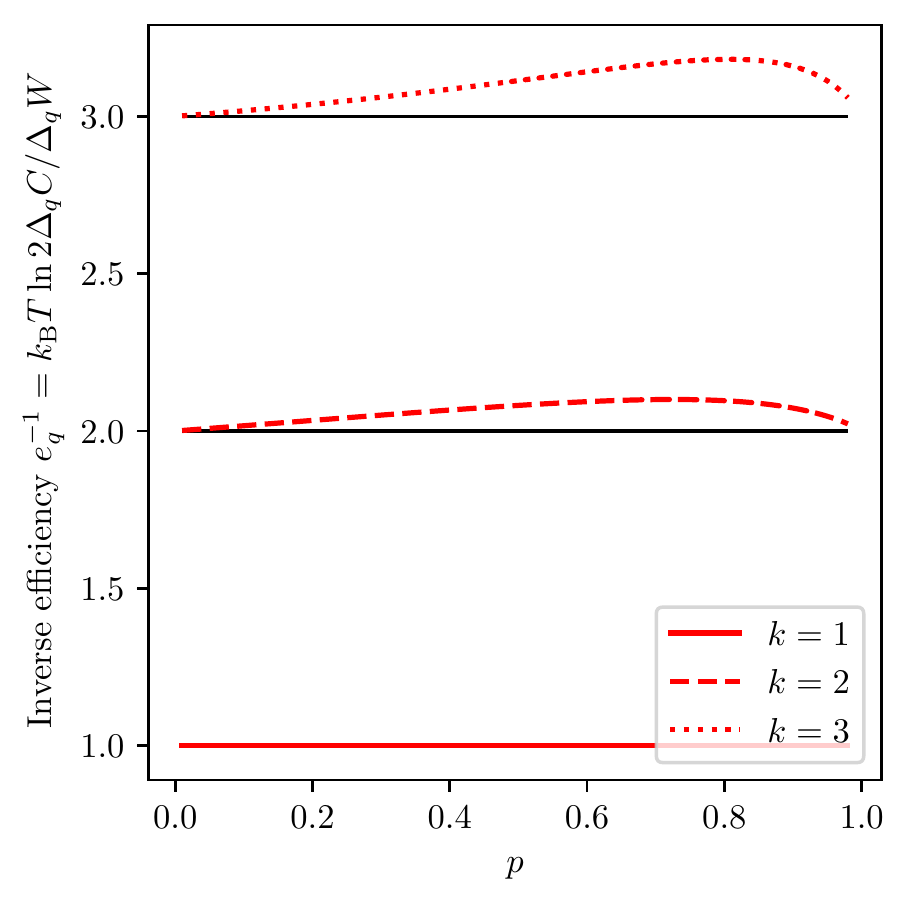}
\caption{$R,k$-Golden Mean Generator thermal efficiency: Inverse compression
	efficiency $e_q^{-1}$ depends only on the crypticity $k$ and transition
	parameter $p$. Black lines added at integer $k$ for comparison.
	}
\label{fig:RKgm}
\end{figure}

There is a family of generators---\emph{$R,k$-Golden Mean Generators}---that
for each integer pair, $R$ and $k$, contains a family that generates processes
with Markov order $R$ and cryptic order $k$, parametrized by a transition
probability $p$. (This family is defined in the SM.) Additionally, for each
$R,k$-Golden Mean Generator the SM shows that (i) the quantum generators are
degenerate and each $\{\phi_{x,k}\}$ gives the same quantum generator and (ii)
the compression efficiency $e_q(R,k)$ of the quantum generator depends only on
the cryptic order $k$. Numerical calculations for $k=1,2,3$ are shown in Fig.
\ref{fig:RKgm}. Note, there, the apparent crypticity bound:
\begin{align}
e_q(k) \leq \frac{1}{k}
  ~.
\label{eq:eff-bound}
\end{align}

{\bf Nemo Generator.}
Most processes have infinite cryptic and Markov orders \cite{Crut08a}:
$R=\infty$ and $k=\infty$. We explored an example of this, the Nemo generator,
whose state-transition diagram is displayed in the SM. Its behavior differs
from $R,k$-Golden Mean Generators in two key respects. First, whereas each
$R,k$-Golden Mean Generator has only one geometrically distinct quantum
implementation, the Nemo generator's space of work and quantum compression
trade-offs is one-dimensional, parametrized by the phase $\Phi = 2\phi_{0A} +
2\phi_{0C} + \phi_{1C} -\phi_{1A}-\phi_{1B}$. Second, the efficiency bound
Eq. \eqref{eq:eff-bound} clearly does not hold. (See Fig. \ref{fig:examples}(e).) If it did, then $e_q=0$. Instead, numerical exploration shows $e_q$ is bounded away from zero and, incidentally, only varies within a small range, such that $e_q \approx 0.3885\pm 0.025$.

{\bf Two-Step Erase Generator.}
The previous two generators had relatively simple quantum implementations with
either complete degeneracy or only filling out a one-dimensional curve in their
\emph{work-compression} ($W/C$) charts. This is not the generic behavior
of quantum generators. To
illustrate this, we now also examine a generator, termed the \emph{Two-Step
Erase} Generator, whose compression thermodynamics is more qualitatively typical
of the generators we explored. Its
state-transition diagram is also given in the SM.

\begin{figure}[t]
\centering
\includegraphics[width=0.9\linewidth]{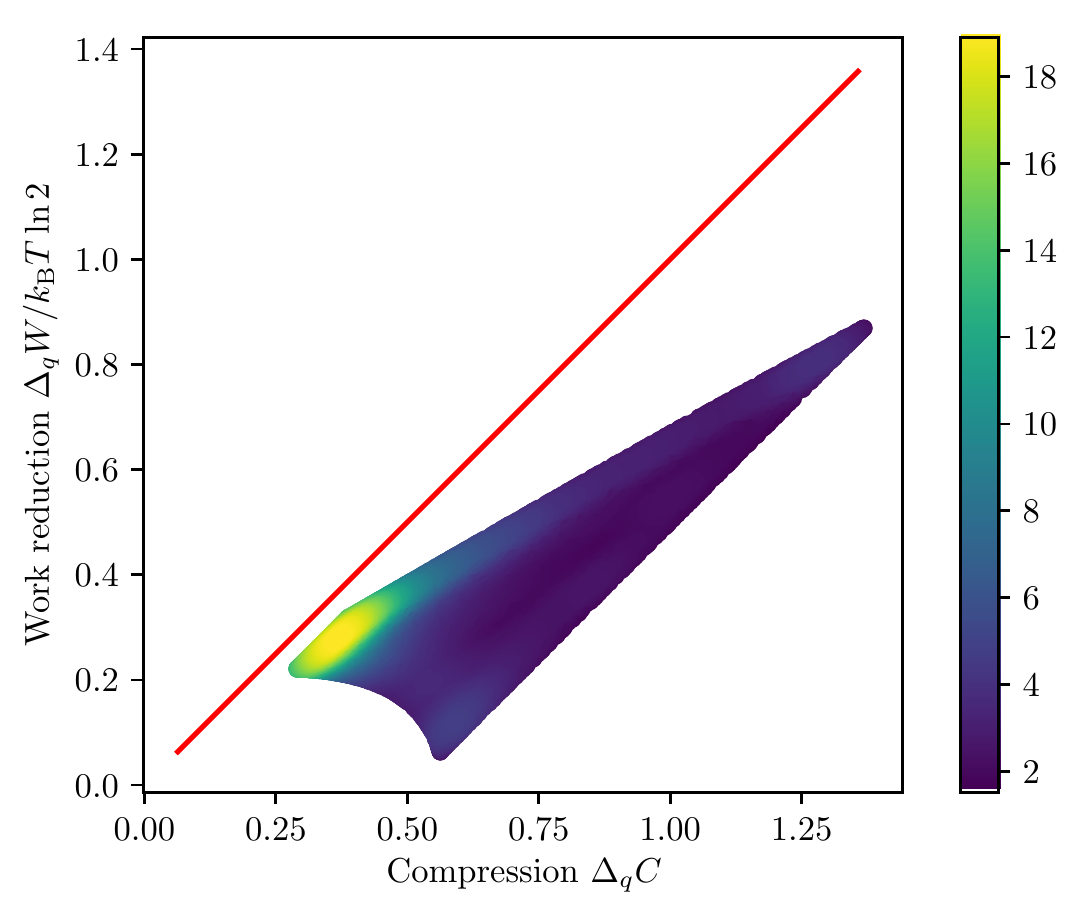}
\caption{Two-Step-Erase Generator: A complex relationship between $\Delta_q W$
	and $\Delta_q C$ appears that is not captured by a single efficiency
	$\epsilon_q$. The density in the plot assumes uniform distribution over
	phases $\left\{\phi_{xs}\right\}$, with blue indicating low density and
	yellow indicating high density.}
\label{fig:twostep}
\end{figure}

Figure \ref{fig:twostep} presents a $W/C$-chart that plots out every achievable
$(\Delta_q C, \Delta_q W/\kB T)$ pair over the range of possible phases
$\left\{\phi_{xs}\right\}$ that determine the quantum implementation, colored
by density. Density is determined by assuming uniform distribution over the
phases $\left\{\phi_{xs}\right\}$. We note that the ``high-advantage'' regions,
where both $\Delta_q C$ and $\Delta_q W$ are large, are actually spanned by
only a small volume of generators, while regions with lower advantages are
spanned by a large volume, indicating that high advantage may not be robust.

{\bf Closing Remarks.} 
We derived single-shot and asymptotic work costs for quantum generator
implementations that are quasistatically attainable. The first of these results
opens the pathway for single-shot comparisons between classical and quantum
resources in process generation, while the second allows direct comparison in
terms of asymptotic quantities. We demonstrated that, when it comes to
quantizing predictive generators, one can ``have their cake and eat it too''
with regards to thermodynamics and memory compression: advantage in both
($\Delta_q C\geq 0$ and $\Delta_q W \geq 0$) can be simultaneously attained.

\vspace{-0.05in}
We analyzed asymptotic thermal efficiencies in four generator classes,
demonstrating a diversity of trade-offs between work and memory advantages. For
every predictive generator of a process, there is a well-defined family of
quantum implementations. However, the scope of their variety ranges from simple
($R,k$-Golden Mean) to highly complex (Two-Step Erase) generators. Even when
the family of generators is simple, there exist fascinating and complex
relationships between the work advantage, memory advantage, and the generator's
computational properties.

\vspace{-0.05in}
Forthcoming work employs the work bounds here to compare quantum \eMs to
classical and quantum nonpredictive generators, seeking the conditions for
optimal work cost over all generators of a given stochastic process
\cite{Loom19b}. It also remains to relax certain assumptions underlying our
work, such as the free availability of empty tapes (which must be prepared),
and the quasistatic assumption which requires infinite time. We believe that
the work completed here provides a foundation for these and other further
extensions.

\vspace{-0.07in}
\paragraph*{Acknowledgments.}
\label{sec:acknowledgments}
The authors thank Ryan James and Fabio Anza for helpful discussions. JPC thanks
the Santa Fe Institute and he and SPL thank the Telluride Science Research
Center for their hospitality during visits. This material is based upon work
supported by, or in part by, FQXi Grant number FQXi-RFP-IPW-1902, the U.S. Army
Research Laboratory and the U. S. Army Research Office under contract
W911NF-13-1-0390 and grant W911NF-18-1-0028, and via Intel Corporation support
of CSC as an Intel Parallel Computing Center.

\vspace{-0.25in}


\onecolumngrid
\clearpage
\begin{center}
\large{Supplementary Materials}\\
\vspace{0.1in}
\emph{\ourTitle}\\
\vspace{0.1in}
{\small
Samuel P. Loomis and James P. Crutchfield
}
\end{center}

\setcounter{equation}{0}
\setcounter{figure}{0}
\setcounter{table}{0}
\setcounter{page}{1}
\makeatletter
\renewcommand{\theequation}{S\arabic{equation}}
\renewcommand{\thefigure}{S\arabic{figure}}
\renewcommand{\thetable}{S\arabic{table}}

The Supplementary Materials calls out energy flow directionality, reviews
quantum implementations of classical generators and the thermodynamic cost of
these implementations, and provides details on the example calculations.

\section{Energy Flow Convention}
\label{sm:FlowDirection}

The main text appeals to a particular direction of energy flow. This is
particularly at issue in applying the information processing Second Law (IPSL)
from Ref. \cite{Boyd15a}. There, the IPSL is stated in the form:
\begin{align*}
W \leq \kB T (\hmu^\prime - \hmu)
  ~,
\end{align*}
where the former entropy rate is that of the output tape and the latter, that
of the input tape. In short, this is in the case in which the ``work done'' $W$
is interpreted as the ``work the ratchet does on a work reservoir''---the work
extracted.

Here, work is defined as the ``work done on the tape, taken from the work
reservoir'', so is opposite in sign. Then one has:
\begin{align*}
W \geq \kB T (\hmu - \hmu^\prime)
  ~.
\end{align*}
And so, the rules of energy flow are consistent, appearing here just with
opposite direction. With simulators, as here, $\hmu = 0$ and therefore work is
extractable while generating a process.

\section{Quantum Implementations of Classical Generators}
\label{sm:QuantumImplment}

We define a classical generator as a triplet $(\mathcal{S}, \mathcal{X},
\left\{ \mathbf{T}^{(x)}:x\in\mathcal{X}\right\})$ where $T^{(x)}_{s's} =
\Pr(s',x|s)$. $\mathcal{S}$ is the finite set of memory states, $\mathcal{X}$
is the finite alphabet of produced symbols, and $\Pr(s',x|s)$ determines the
transition-and-production dynamic of the generator.

To analyze the thermodynamics of physical implementations of generators, we must establish
rules that circumscribe what we consider physically allowed and the
correspondence to thermodynamic quantities such as work and heat.

Here, we consider the \emph{resource theory of thermal
operations} \cite{Horo13a,Bran13a}. Generally, on a quantum system $S$ we allow
operations of the form:
\begin{align}
\mathcal{E}\left(\rho_S\right) := \mathrm{Tr}_B\left(U\rho_S \otimes \frac{e^{-\beta H_B}}{Z_B} U^\dagger\right)
  ~,
\label{eq:thermal}
\end{align}
where $S$ and $B$ are auxiliary systems with Hamiltonians $H_S$ and $H_B$, $B$
a thermal bath, and $U$ acts on the joint Hilbert space of $\mathcal{H}_S$ and
$\mathcal{H}_B$. The unitary operator $U$ satisfies the rule of
microscopic energy conservation, where we constrain $[U,H_S+H_B]=0$.

Recall from the main body that an implementation
$\left(\mathcal{H}_S,\mathcal{H}_X,\mathcal{H}_A,\mathcal{H}_B, U,
\mathcal{E}\right)$ of a generator involves the memory space $\mathcal{H}_S$,
symbol space $\mathcal{H}_X$, auxiliary space $\mathcal{H}_A$, and bath space
$\mathcal{H}_B$; the ensemble $\mathcal{E}=\left\{
\ket{\psi_s}:s\in\mathcal{S}\right\}$; and a unitary acting on
$\mathcal{H}_S\otimes \mathcal{H}_X\otimes \mathcal{H}_A\otimes \mathcal{H}_B$,
such that the channel:
\begin{align*}
\mathcal{T}_{SX}\left(\rho_{SX}\right) :=\mathrm{Tr}_{AB} \left(U\rho_{SX}\otimes \ketbra{0}_A \otimes \rho_B U^\dagger\right)
\end{align*}
satisfies:
\begin{align}
\label{channel-def}
\mathcal{T}_{SX} \left(\ketbra{\psi_s}_S\otimes \ketbra{0}_X\right) = 
\sum_{s',x} \Pr\left(s',x|s\right) \ketbra{\psi_{s'}}\otimes \ketbra{x}_X
  ~.
\end{align}
Suppose that there are Hamiltonians $H_S$, $H_X$, $H_A$, and $H_B$ for each
system such that $\rho_\text{sys} = Z_\text{sys}^{-1}\exp\left(-\beta
H_\text{sys} \right)$ is
the Gibbs distribution of its Hamiltonian. Then the resource theory of thermal
operations requires $[U,H_S+H_X+H_A+H_B]=0$ \cite{Janz00a}.

In quantum mechanics, the rule of microscopic energy conservation $[U,H_S+H_X+H_A+H_B]=0$ brings coherence
with respect to the Hamiltonian into play as a resource \cite{Lost15a,Lost18a}.
The type of systems we consider here are what are often, in the literature of
information engines, called \emph{information reservoirs}: systems whose
Hamiltonian is trivially degenerate, so that energetics does not play a direct role
in their dynamics. On such systems, tracking coherence is no longer at issue, as
all operators commute with a degenerate Hamiltonian.

In this case, we can restrict ourselves to Hamiltonian control protocols
for the erasure of the auxiliary system, as in \cite{Rio11a},
where the Hamiltonian is degenerate at the beginning and end of the procedure,
and stays in one basis during the procedure. In this case
the work cost is computed by adding changes in the energy levels, weighted
by the relative probabilities of being in those levels \cite{Vinj16a}.

Among quantum implementations, the only form that has been studied for
generators is the \emph{unitary} implementation, which itself is only valid for
predictive generators \cite{Bind17a,Liu19a}. In this implementation, the bath
system $B$ is not used and the unitary operator is split into two steps,
$U=U_2 U_1$, where $U_1 = U_{SX}\otimes 1_A$ and $U_2 = 1_S\otimes U_{XA}$.

In the first step, the \emph{evolution step}, we act only on the memory and the
output $SX$ with the unitary $U_{SX}$ defined by the action:
\begin{align}
\label{qmach-def}
U_{SX}\ket{\psi_s}\ket{0} = \sum_{x} e^{i\phi_{xs}} \sqrt{\Pr(x|s)} \ket{\psi_{f(x,s)}}\ket{x}
  ~.
\end{align}
The unitarity evolution of $U_{SX}$ in fact defines the overlap matrix
$\Omega_{rs} := \braket{\psi_r | \psi_s}$ via the recursive formula
\cite{Liu19a}:
\begin{align}
\label{overlap-constraint}
\Omega_{rs} = \sum_x \sqrt{\Pr(x|r)\Pr(x|s)} e^{i(\phi_{xs}-\phi_{xr})} \Omega_{f(r,s) f(x,s)}
  ~.
\end{align}
For any choice of phases, an overlap matrix $\Omega_{rs}$ and unitary $U_{SX}$
exist.

In the second step---the \emph{measurement step}---the symbol is observed,
sending the pure state $U_{SX} \ketbra{\psi_s}\otimes \ketbra{0}
U_{SX}^\dagger$ to the mixed state in Eq. \eqref{channel-def}. This is done by
coupling the system $X$ to the auxiliary system $A$ and applying a unitary so
that:
\begin{align*}
U_{XA}\ket{x}_X\ket{0}_A  \propto \ket{x}_X\ket{x}_A
  ~.
\end{align*}
When the auxiliary is discarded (or, more realistically, reset) we are left with the state on $SX$, as desired.

The perfect preparation, or reset, of the auxiliary in state $\ket{0}_A$ cannot
be performed with finite resources \cite{Gury20a}. In the following section we
consider the work costs achievable in the single-shot setting where a
probability of failure $\epsilon$ is allowed for the reset step. The underlying
protocol is quasistatic (see Ref. \cite{Rio11a}), requiring infinite time to
complete. Here, we focus on fundamental work limits in the single-shot and
asymptotically parallel regimes, but not in the finite time regime, which is an
important direction of future extension.

\section{Information-theoretic definitions}
\label{sm:InfoTheory}
This section employs the Shannon \cite{Cove06a}, von Neumann \cite{Niel10a}, 
and smooth conditional entropies \cite{Renn04a,Renn05a}. 
The Shannon entropy for a random variable $X\sim \Pr(x)$ 
and von Neumann entropy for a system $A$ (no relation to the auxiliary $A$ in the previous section) 
with density matrix $\rho_A$ are, respectively:
\begin{align*}
        \H{X} &\equiv -\sum_x \Pr(x)\log_2 \Pr(x) ~ \text{and}\\
        \Hq{S} &\equiv -\mathrm{Tr}\left(\rho_S\log_2 \rho_S\right)
  ~.
\end{align*}
For bipartite variables $XY$ and bipartite quantum systems $AB$, these quantities beget
the conditional entropies and mutual informations:
\begin{align*}
        \H{X|Y} &\equiv \H{XY}-\H{Y} ~,\\
        \Hq{A|B} &\equiv \Hq{AB}-\Hq{B} ~,\\
        \mathrm{I}[X:Y] &\equiv \H{X} +\H{Y}-\H{XY} ~,~ \text{and}\\
        \mathrm{I}_q[A:B] &\equiv \Hq{A} +\Hq{B}-\Hq{AB}
  ~.
\end{align*}
For two systems $A$ and $B$ with joint state $\rho_{AB}$, the min- and
max-entropies are given by \cite{Renn04a,Renn05a,Toma13a,Toma2016}:
\begin{align*}
        \Hmin{}{A|B}_\rho &\equiv \min_{\sigma_B} \sup \{\lambda:\rho_{AB}\leq 2^{-\lambda}1_A\otimes \sigma_B\} ~ \text{and}\\
        \Hmax{}{A|B}_\rho &\equiv \max_{\sigma_B} 2\log_2 F\left(\rho_{AB},1_A\otimes \sigma_B\right)
  ~,
\end{align*}
where $F(\rho,\sigma) =
\mathrm{Tr}\left(\sqrt{\sqrt{\rho}\sigma\sqrt{\rho}}\right)$ is the
\emph{fidelity}. The smooth conditional entropies are optimizations of these quantities
over all $\tilde{\rho}{}_{AB}$ within the $\epsilon$-ball
$B_\epsilon(\rho_{AB})$; that is, all states such that
$\sqrt{1-F\left(\tilde{\rho}{}_{AB},\rho_{AB}\right)}<\epsilon$:
\begin{align*}
        \Hmin{\epsilon}{A|B} &\equiv \max_{\tilde{\rho}{}_{AB}}\Hmin{}{A|B}_{\tilde{\rho}} ~ \text{and}\\
        \Hmax{\epsilon}{A|B} &\equiv  \min_{\tilde{\rho}{}_{AB}}\Hmax{}{A|B}_{\tilde{\rho}}
  ~.
\end{align*}
When $B$ is uncorrelated with $A$, $\rho_{AB} = \rho_A\otimes \rho_B$, the
resulting quantities are independent of $B$ and so we have the marginal smooth
entropies $\Hmax{\epsilon}{A}$ and $\Hmin{\epsilon}{A}$. We will utilize a
result on smooth conditional entropies that generalizes the chain rule on von
Neumann entropy \cite{Vita13a}. We state two somewhat streamlined versions of
the theorem here. For any $\delta>0$ and systems $AB$:\footnote{Comparing our
statements with Ref. \cite{Vita13a}, note that ignore the third system $C$
and swap $A$ and $B$. Rather than use the four parameters $\epsilon$,
$\epsilon'$, $\epsilon''$, and $f$, we use the single parameter $\delta$ such
that $\epsilon = 4\delta$, $\epsilon' = \epsilon''=\delta$, and $f =
O\left(\log \frac{1}{\delta}\right)$. Then Eq. \eqref{chain-rule-1} and Eq.
\eqref{chain-rule-2} correspond to the sixth and first equations on page 2 of
Ref. \cite{Vita13a}.}
\begin{align}
\label{chain-rule-1}
\Hmax{\delta}{B\middle|A}&\leq 
\Hmax{4\delta}{AB} - \Hmin{\delta}{A}
{\color{mygray} +O\left(\log\frac{1}{\delta}\right)} ~ \text{and}\\
\label{chain-rule-2}
\Hmin{\delta}{B\middle|A} &\leq 
\Hmin{4\delta}{AB} - \Hmin{\delta}{A}
{\color{mygray} +O\left(\log\frac{1}{\delta}\right)}
\end{align}

\section{Generator implementation costs}
\label{sm:ImplementCost}

We import the following result from Ref. \cite{Rio11a}: Given a system $S$
correlated with an auxiliary $A$, and any $\epsilon>0$, there is a procedure
for erasing $A$ while preserving $S$, with probability of failure $\epsilon$,
that has a work cost of no more than:\footnote{Again, comparing with Ref.
\cite{Rio11a}, our Eq. \eqref{single-shot-erase} is drawn from Thm. 1. System
$S$ from Ref. \cite{Rio11a} is system $A$ here, system $O$ from Ref.
\cite{Rio11a} is system $S$ here, and $\delta$ from Ref. \cite{Rio11a} is
$\epsilon$ here. In Ref. \cite{Rio11a}, take $\epsilon = \delta^2/13$ and
$\Delta = 2\log\left(1/\epsilon\right) = 2\log 13 + 4\log
\left(1/\delta\right)$. The term $2\log 13$ is irrelevant in the $\delta
\rightarrow 0$ limit so we include it in the big-O term.}
\begin{align}
\label{single-shot-erase}
\frac{W}{\kB T\ln 2}\leq \Hmax{\epsilon^2/16}{A\middle|S}
{\color{mygray} + O\left(\log\frac{1}{\epsilon}\right)}
  ~.
\end{align}
The smooth conditional entropy also provides a lower bound on the attainable
work cost\footnote{In this case, comparing to Ref. \cite{Fais15a}, note that
their system $E$ is our system $A$, their system $X'$ is our system $S$, and
finally we have directly replaced $\bar{\epsilon}$ with $\sqrt{2\epsilon}$.}
\cite{Fais15a}: $W/\kB T\ln 2 \geq \Hmax{\sqrt{2\epsilon}}{A\middle|S}$. While
this bound is finite in the limit $\epsilon\rightarrow 0$, the upper bound Eq.
(\ref{single-shot-erase}) diverges and so does not guarantee any particular
attainable work cost. Additionally, it must be noted that the protocol used by
Ref. \cite{Rio11a} is quasistatic, requiring infinite time to complete. This is
consistent with previous results on, say, the unattainability of perfect
measurements with finite resources \cite{Gury20a}.

We can use Eq. (\ref{single-shot-erase}) to prove a generalization of the
\emph{detailed} Landauer cost---that is, to show the conditions under which
Landauer's bound is quasistatically attainable. Suppose we have a quantum
channel $\mathcal{E}$ we wish to implement and we do so on a system $S$ with
mixed state $\rho_S$. The target state is $\rho_{S}' =
\mathcal{E}\left(\rho_S\right)$. We perform the map in the following way.
Using the Stinespring dilation of $\mathcal{E}$, we couple $S$ to an auxiliary
system $A$ in state $\ketbra{0}_A$ and perform a unitary operation on both
systems:
\begin{align*}
\rho_{SA}' = U_{SA} \rho_S\otimes \ketbra{0}_A U_{SA}^{\dagger}
  ~,
\end{align*}
such that $\mathcal{E}\left(\rho_S\right) =
\mathrm{Tr}_A\left(\rho_{SA}'\right)$. At the end of the procedure we must
erase $A$. This can be done with cost Eq. \eqref{single-shot-erase}. This form
of the cost for implementing a channel is given in Ref. \cite{Fais15a}, where
an argument similar to the following was used to derive Landauer's lower bound
in the macroscopic limit. Here, we apply the same logic to show that Landauer's
bound is also \emph{attainable} in the macroscopic (simulating an infinite
number of parallel channels) limit.

Applying Eq. \eqref{chain-rule-1} to Eq. \eqref{single-shot-erase}, we have:
\begin{align*}
\frac{W}{\kB T\ln 2} \leq \Hmax{\epsilon^2/4}{S'A'}-\Hmin{\epsilon^2/16}{S'}
{\color{mygray} + O\left(\log\frac{1}{\epsilon}\right)}
  ~.
\end{align*}
However, $\Hmax{\epsilon^2/4}{S'A'} = \Hmax{\epsilon^2/4}{S}$ by unitary
equivalence, so we have the erasure cost:
\begin{align}
\label{single-shot-landauer}
\frac{W}{\kB T\ln 2} \leq \Hmax{\epsilon^2/4}{S}-\Hmin{\epsilon^2/16}{S'}
{\color{mygray} + O\left(\log\frac{1}{\epsilon}\right)}
  ~.
\end{align}
Since we can perform the initial unitary with no work, this is the only work
cost involved in implementing the channel. To summarize: The channel
$\mathcal{E}$ can be performed on the system $S$ with a work cost not exceeding
Eq. \eqref{single-shot-landauer}.

Now, suppose we choose instead to implement parallel generation of our process.
That is, we have $N$ independent systems on which we want to implement $N$
independent copies of the channel $\mathcal{E}$ with probability of error less
than $\epsilon>0$. Naturally, the work cost becomes:
\begin{align*}
        \frac{W}{\kB T\ln 2} \leq \Hmax{\epsilon^2/4}{S^{\otimes N}}-\Hmin{\epsilon^2/16}{S'^{\otimes N}}
        {\color{mygray} + O\left(\log\frac{1}{\epsilon}\right)}
  ~.
\end{align*}
Significantly, the error term does not depend on $N$. When we further account
for the Asymptotic Equipartition Theorem of smooth conditional entropies, we have the
remarkable result for the work rate:
\begin{align}
\label{landauer-asymptotic}
        \frac{W}{N \kB T\ln 2} \leq \H{S}-\H{S'}
        {\color{mygray} + O\left(\sqrt{\frac{1}{N}\log\frac{1}{\epsilon}}\right)}
  ~.
\end{align}
With Landauer's bound sandwiching the work from below, we find a tight result
on the achievable work cost. By scaling error with $N$, for instance $\epsilon
\sim 2^{-\sqrt{N}}$, Landauer's bound can, in the limit $N\rightarrow \infty$,
be achieved for quantum channels. In the single-shot regime, the bound of Eq.
\eqref{single-shot-landauer} gives us a somewhat less certain range of
achievability.

This can be applied directly to the implementation of {\em generators}
discussed in the previous section. In the single-shot setting, we have:
\begin{align*}
        \frac{W}{\kB T\ln 2} \leq \Hmax{\epsilon^2/4}{S}-\Hmin{\epsilon^2/16}{S'X}
        {\color{mygray} + O\left(\log\frac{1}{\epsilon}\right)}
  ~.
\end{align*}
Applying Eq. \eqref{chain-rule-2} this becomes:
\begin{align}
\label{single-shot-modularity}
        \frac{W}{\kB T\ln 2} \leq \ \Hmax{\epsilon^2/4}{S}-\Hmin{\epsilon^2/64}{S'\middle|X}
        - \Hmin{\epsilon^2/64}{X}
        {\color{mygray} + O\left(\log\frac{1}{\epsilon}\right)}
  ~.
\end{align}
Finally, consider the asymptotic limit of $N$ parallel generators producing $N$
independent copies of a stochastic process. The Asymptotic Equipartition
Theorem again gives the result:
\begin{align}
\label{asymptotic-modularity}
        \frac{W_q}{\kB T\ln 2} \leq \ \Hq{S}-\Hq{S'\middle|X}
        - \H{X}{\color{mygray} + O\left(\sqrt{\frac{1}{N}\log\frac{1}{\epsilon}}\right)}
  ~.
\end{align}

Our last result is the inequality Eq. \eqref{inequality}. We note that:
\begin{align*}
W_q & = \Hq{S}-\Hq{S'\middle|X} - \H{X} \\
  & = \Hq{S'}-\Hq{S'\middle|X} - \H{X} \\
  & = \mathrm{I}_q\left[S':X\right] -\H{X}
  ~.
\end{align*}
The same form can be given for $W_\mu$ in terms of the Shannon entropies:
\begin{align*}
        W_\mu & = \H{S}-\H{S'\middle|X} - \H{X} \\
        & = \mathrm{I}\left[S':X\right] - \H{X}
  ~.
\end{align*}
Now, in the quantum model $\mathrm{I}_q\left({S':X}\right)$ is the mutual information of the
state:
\begin{align*}
        \rho_{SX}' = \sum_{x,s,s'} \Pr{}_0(s)\Pr(s',x|s)\ketbra{\psi_{s'}}\otimes\ketbra{x}
  ~,
\end{align*}
which can be derived from the classical variables $S'X$ by the local mappings
$s'\mapsto \ket{\psi_{s'}}$ and $x\mapsto \ket{x}$. Then by the data processing
inequality: $\mathrm{I}_q\left(S':X\right) \leq \mathrm{I}\left(S':X\right)$.
This proves that $W_q \leq W_\mu$.

\begin{figure}
\begin{tabular}{ccc}
\includegraphics[height=0.3\textwidth]{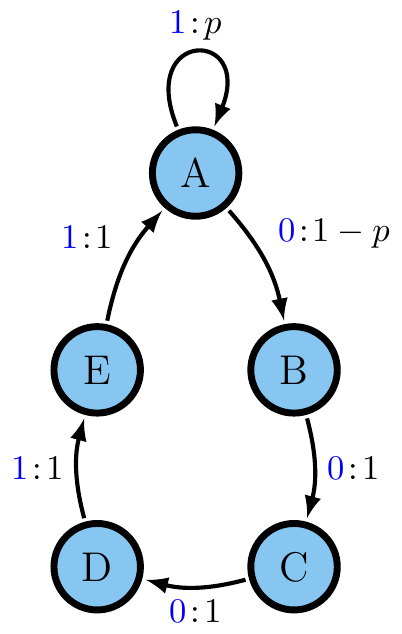} &
\includegraphics[width=0.25\textwidth]{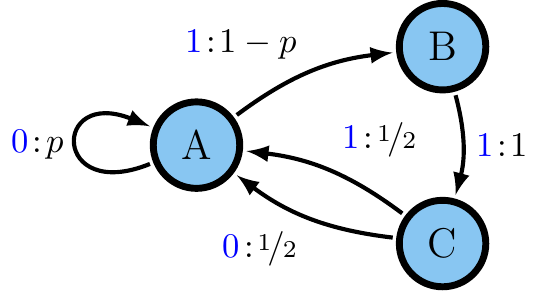}&
\includegraphics[width=0.3\textwidth]{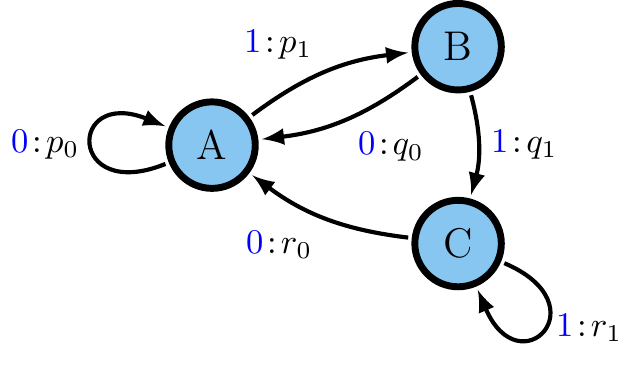} \\
(a) $3,2$-Golden Mean & (b) Nemo & (c) Two-Step Erase \\[6pt]
\includegraphics[width=0.3\textwidth]{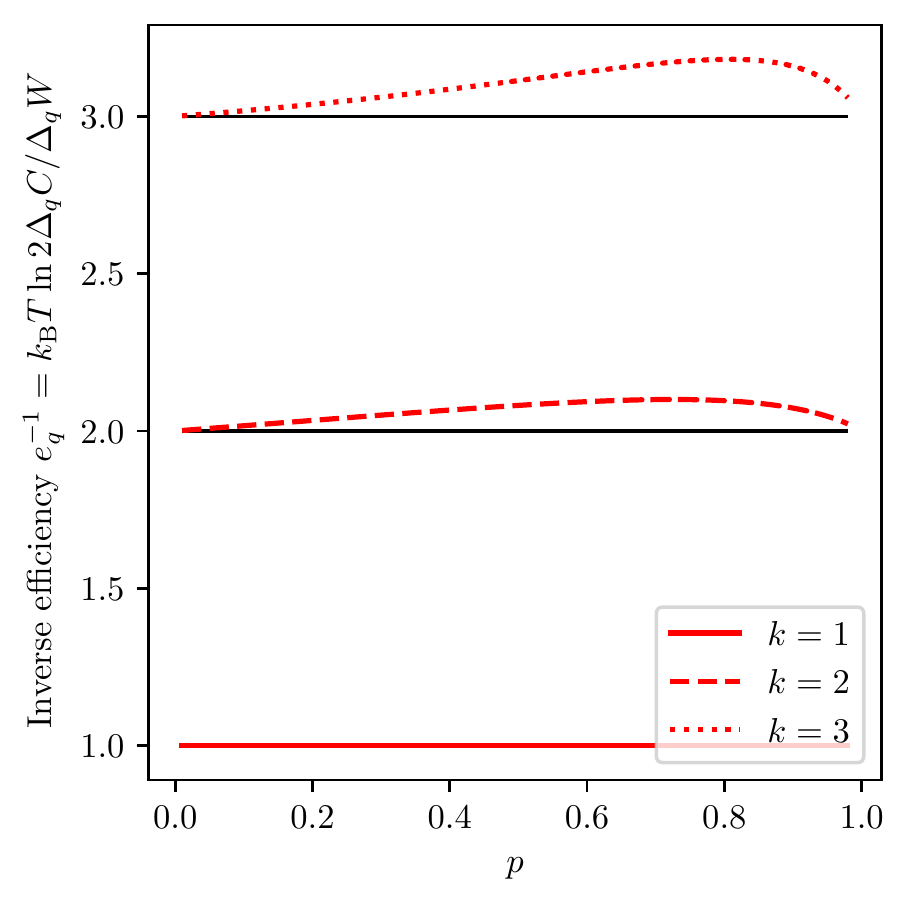} &
\includegraphics[width=0.315\textwidth]{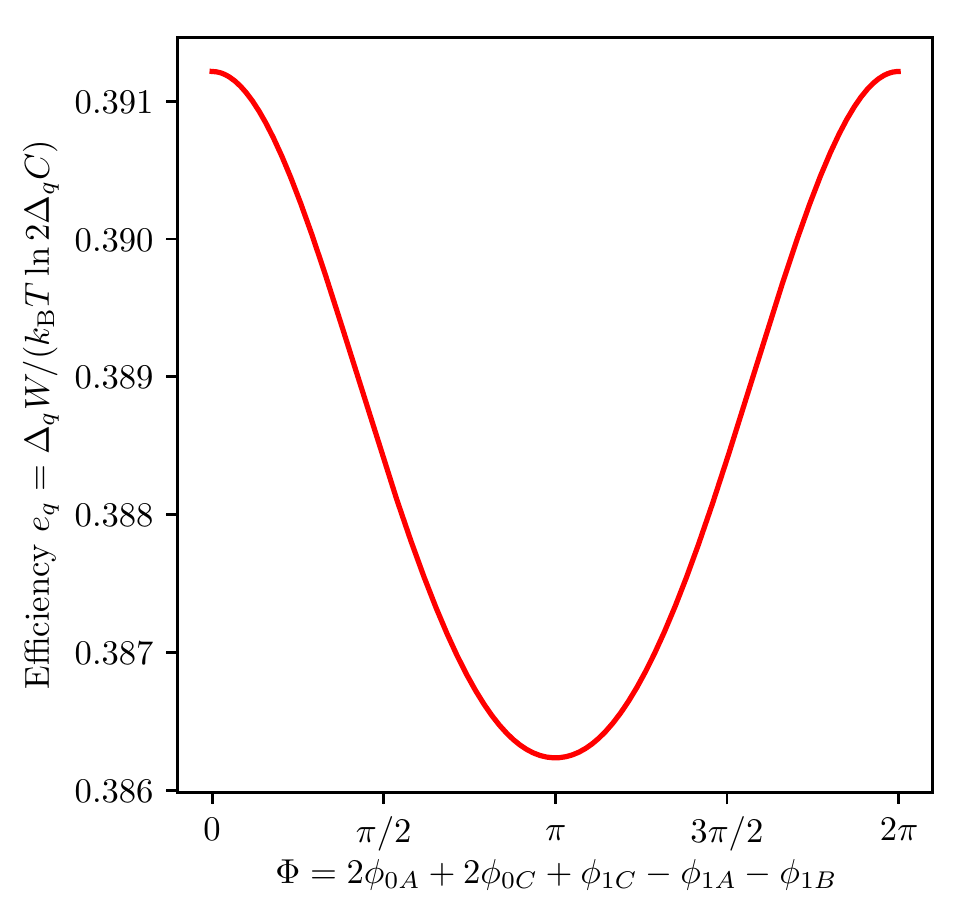} &
\includegraphics[width=0.35\textwidth]{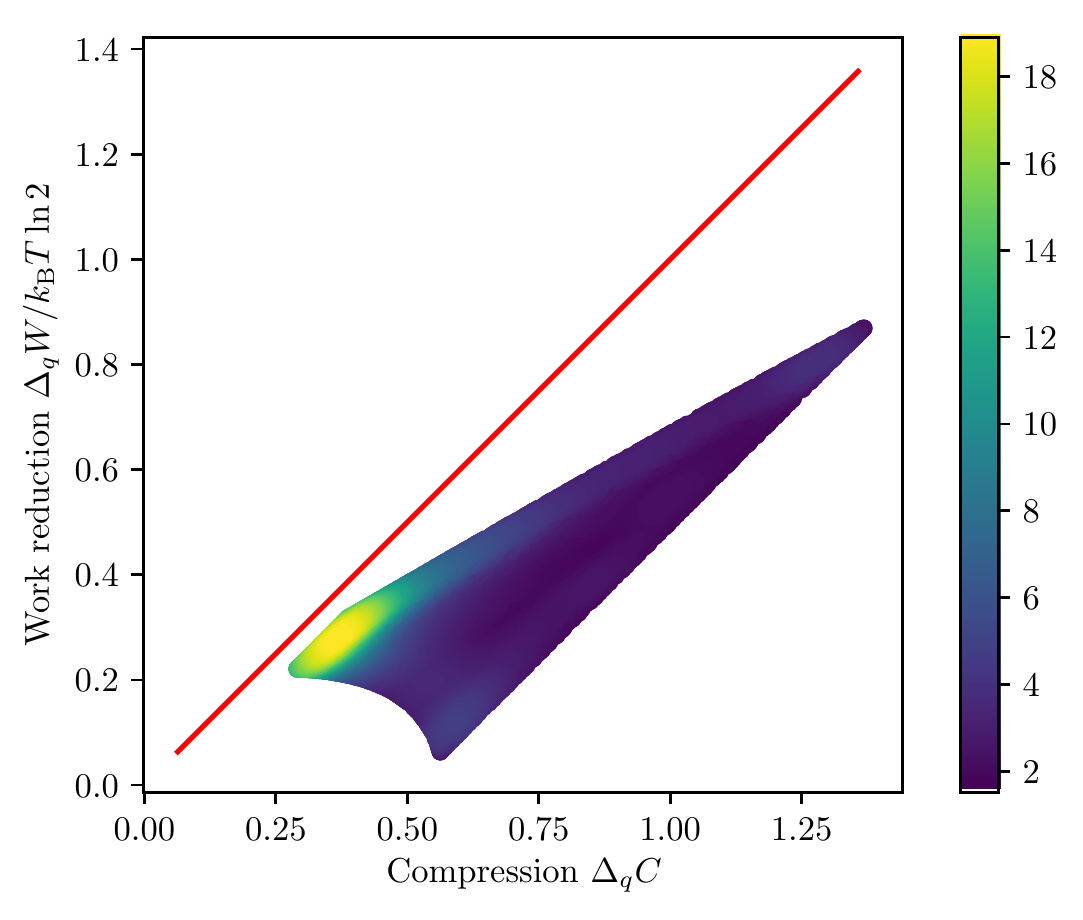} \\
(d)  & (e) & (f)
\end{tabular}
\caption{\emph{(a)} The $R,k$-Golden Mean Generator for $R=3$ and $k=2$. See
	main text for general construction. \emph{(b)} The Nemo generator, which
	has infinite Markov and cryptic orders. \emph{(c)} The Two-Step Erase
	generator always erases its memory to $A$ upon generating a $0$, but erases
	its memory to $C$ upon generating two consecutive $1$'s. \emph{(d)} The
	inverse compression efficiency $e_q^{-1}$ of the $R,k$-Golden Mean
	generator depends only on the crypticity $k$ and transition parameter $p$.
	Black bars added at integer values for comparison. \emph{(e)} For the Nemo
	generator, the overlap matrix $\Omega$, which determines all the
	quantum-information-theoretic properties of the implementation, depends
	only on the combined phase $\Phi =
	2\phi_{0A}+2\phi_{0C}+\phi_{1C}-\phi_{1A}-\phi_{1B}$. Consequently, the
	efficiency $e_q$ depends on this quantity and the parameter $p$. Numerical
	exploration shows that the variation of $e_q$ due to $\Phi$ is quite small
	in amplitude and varies sinusoidally. We plot this variation for $p=0.5$.
	\emph{(f)} For general $p,q,r$ of the Two-Step Erase process (showing
	$p=1/2$, $q=1/5$, and $r=2/5$ here), we find a complex relationship between
	$\Delta_q W$ and $\Delta_q C$ appears. This is not captured by a single
	efficiency $e_q$. The density in the plot assumes uniform distribution over
	phases $\left\{\phi_{xs}\right\}$, with blue indicating low density and
	yellow indicating high density.
	}
\label{fig:examples}
\end{figure}

\section{Example generators}
\label{sm:ExampleCalculations}

Understanding the behavior of our example generators (see Fig.
\ref{fig:examples}) requires discussing gauge properties of quantum
implementations.

The physical properties of each quantum generator are entirely determined by
its overlap matrix $\Omega_{rs} = \braket{\psi_r | \psi_s}$. However, this in
itself contains nonphysical degrees of freedom \cite{Loom18a}. None of the
invariant geometry of our generators is modified under the transformation
$\ket{\psi_s} \mapsto e^{i\Psi_s} \ket{\psi_s}$ on the signal states. Thus,
these represent a \emph{gauge transformation}. In terms of the overlap matrix,
this means that our generators are invariant under the transformations
$\Omega_{rs} \mapsto e^{i(\Psi_s - \Psi_r)}\Omega_{rs}$.

It is helpful (especially for the Nemo process) to consider these gauge
properties in terms of how they act on the phases $\{\phi_{xs}\}$ that
determine the quantum generator. Applying the gauge transformation to the
consistency formula gives:
\begin{align*}
\Omega_{rs} = \sum_x \sqrt{\Pr(x|r)\Pr(x|s)}
  e^{i(\tilde{\phi}_{xs}-\tilde{\phi}_{xr})} \Omega_{f(x,r) f(x,s)}
  ~,
\end{align*}
where:
\begin{align}
\label{induced}
        \tilde{\phi}_{xs}=\phi_{xs}-\Psi_s+\Psi_{f(x,s)}
\end{align}
is the induced transformation on the generator's phases. Equation
\eqref{induced} can be taken as a fundamental description of the gauge
transformation.

Using Eq. \eqref{induced} allows us to determine the \emph{gauge
invariants}---that is, combinations of the phases $\{\phi_{xs}\}$ that do not
change under a gauge transformation. In this case, the gauge invariants are
best understood graphically, in terms of the hidden Markov models from before.
Each phase $\{\phi_{xs}\}$ can be understood as being assigned to an edge,
while each phase in the gauge transformation $\{\Psi_s\}$ can be seen as being
assigned to a state. 

For each loop of edges, we can take a linear combination of the constituent
edges' phases $\phi_{xs}$, adding positive and negative signs based on the
direction of the edges. These loop sums are the gauge invariants. For instance,
the Nemo process has $\Phi_0 = \phi_{0A}$, $\Phi_1 = \phi_{1C}-\phi_{0C}$, and
$\Phi_{2}=\phi_{1A}+\phi_{1B}+\phi_{1C}$ as gauge invariants.

\subsection{$(R,k)$-Golden Mean Generators}

An $R,k$-Golden Mean Generator is one with $R+k$ memory states. These states
can be considered to belong to two groups: the $A$ state, which is the only
nondeterministic state and the $B$-states $\mathcal{B} \equiv
\{B_1,\dots,B_{R+k-1}\}$. The $B$-states are further broken down into a Markov
part $\mathcal{R}\equiv\{B_1,\dots,B_{R-1}\}$ and a cryptic part
$\mathcal{K}\equiv \{B_{R},\dots,B_{R+k-1}\}$. The dynamic on the generator is given by:
\begin{align*}
\Pr\left(s',0\middle|s\right) & = \begin{cases}
1-p & s=A,\ s'=B_1\\
1 & s=B_r,\ s'=B_{r+1},\ 0\leq r< R\\
0 & \mathrm{otherwise}
\end{cases}\\
\text{and}\\
\Pr\left(s',1\middle|s\right) & = \begin{cases}
p & s',s=A\\
1  & s=B_r,\ s'=B_{r+1},\ R\leq r\leq R+k-2\\
1 & s=B_{R+k-1},\ s'=A\\
0 & \mathrm{otherwise}
\end{cases}
  ~.
\end{align*}
We can check that:
\begin{align*}
\Pr{}_0(s) = \begin{cases}
\frac{1}{1+(R+k-1)(1-p)} & s=A\\
\frac{1-p}{1+(R+k-1)(1-p)} & s=A
\end{cases}
  ~.
\end{align*}
is the stationary distribution. Letting $Z=1+(R+k-1)(1-p)$, we have:
\begin{align*}
\Pr\left(s',0,s\right) & = \begin{cases}
\frac{1-p}{Z} & s=A,\ s'=B_1\\
\frac{1-p}{Z} & s=B_r,\ s'=B_{r+1},\ 0\leq r< R\\
0 & \mathrm{otherwise}
\end{cases} \\
\text{and}\\
\Pr\left(s',1,s\right) & = \begin{cases}
\frac{p}{Z} & s',s=A\\
\frac{1-p}{Z} & s=B_r,\ s'=B_{r+1},\  R\leq r\leq R+k-2\\
\frac{1-p}{Z} & s=B_{R+k-1},\ s'=A\\
0 & \mathrm{otherwise}
\end{cases}
  ~.
\end{align*}
It is helpful to also have:
\begin{align*}
\Pr(X=0) & = \frac{R(1-p)}{Z} ~,\\
\Pr(X=1) & = \frac{(k-1)(1-p)-1}{Z} ~,\\
\Pr\left(s'\middle|0\right) & = \begin{cases}
\frac{1}{R} & s=A,\ s'=B_1\\
\frac{1}{R} & s=B_r,\ s'=B_{r+1},\ 0\leq r< R\\
0 & \mathrm{otherwise,}
\end{cases}\\
\text{and}\\
\Pr\left(s'\middle|1\right) & = \begin{cases}
\frac{p}{(k-1)(1-p)-1} & s',s=A\\
\frac{1-p}{(k-1)(1-p)-1} & s=B_r,\ s'=B_{r+1},\  R\leq r\leq R+k-2\\
\frac{1-p}{(k-1)(1-p)-1} & s=B_{R+k-1},\ s'=A\\
0 & \mathrm{otherwise}
\end{cases}
  ~.
\end{align*}

First, we wish to show that regardless of the chosen phases $\{\phi_{xs}\}$ we
get the equivalent quantum model. Recall that the formula defining the overlaps
is given by:
\begin{align*}
\Omega_{rs} = \sum_x \sqrt{\Pr(x|r)\Pr(x|s)} e^{i(\phi_{xs}-\phi_{xr})} \Omega_{f(r,s) f(x,s)}
  ~.
\end{align*}
In this case, we have:
\begin{align*}
\Omega_{AB_{R+k-1}} & =\sqrt{p} e^{i(\phi_{1B_{R+k-1}}-\phi_{1A})} ~,\\
\Omega_{B_{r}B_{s}} & = e^{i(\phi_{1B_{r}}-\phi_{1B_s})}\Omega_{B_{r+1}B_{s+1}}
~\text{and}\\
\Omega_{AB_{r}} & =\sqrt{p}e^{i(\phi_{1B_{r}}-\phi_{1A})}\Omega_{AB_{r+1}}
  ~,
\end{align*}
which has the solution:
\begin{align*}
\frac{\Omega_{AB_{R+m}}}{\sqrt{p^{k-m}}}
  & = e^{i\left(\sum_{j=m}^{k-1}\phi_{1B_{R+j}}-(k-m)\phi_{1A}\right)}
  ~\text{and}\\
\frac{\Omega_{B_{R+m}B_{R+n}}}{\sqrt{p^{m-n}}}
  & =  e^{i\left(\sum_{j=n}^{k-1}\phi_{1B_{R+j}}-\sum_{j=m}^{k-1}\phi_{1B_{R+j}}-(m-n)\phi_{1A}\right)}
  ~.
\end{align*}
Note that under the gauge transformation $\Psi_{A} = k\phi_{1A}$ and
$\Psi_{B_m} = \sum_{j=m}^{k-1}\phi_{1B_{R+j}}+m\phi_{1A}$, we can eliminate
phases and end up simply with:
\begin{gather}
\label{golden-mean-overlap}
\begin{split}
\Omega_{AB_{R+m}} =\sqrt{p^{k-m}}  ~\text{and}\\
\Omega_{B_{R+m}B_{R+n}} = \sqrt{p^{m-n}}
\end{split}
~.
\end{gather}
We note that this matrix only explicitly depends upon $k$ and not $R$. This
extends a result from Ref. \cite{Riec15b} to all $R$ and $k$, as well as to all
choices of phase $\{\phi_{xs}\}$.

We can also apply these probabilities to compute the efficiency.  The
conditional entropies are:
\begin{align*}
\H{S'|X=0} =&\log R ~\text{and}\\
\H{S'|X=1} =& \log \left(k(1-p)+p\right) - \frac{(k-1)(1-p)}{k(1-p)+p}\log(1-p)
  ~.
\end{align*}
Under compression, the $X=0$ term does not change: $\Hq{S'|X=0}=\log R$. We
will not compute the compressed term for $X=1$ since we need only note that it
is a function of $k$ and $p$ and not of $R$.

The classical and quantum memories can be evaluated as:
\begin{align*}
C_\mu =& \log Z + \frac{1}{Z}C_\mu^{(\mathcal{K})} - \frac{1}{Z} (R-1)(1-p)\log
(1-p) ~\text{and}\\
C_q =& \log Z + \frac{1}{Z}C_q^{(\mathcal{K})} - \frac{1}{Z} (R-1)(1-p)\log (1-p)
  ~,
\end{align*}
where:
\begin{align*}
C_\mu^{(\mathcal{K})} =& (k-1) \log (1-p) ~\text{and}\\
C_q^{(\mathcal{K})} =&\mathrm{Tr}\left( \Omega ZP^{(\mathcal{K})}\log\left(\Omega ZP^{(\mathcal{K})}\right)  \right)
  ~.
\end{align*}
are the contributions to complexity from only the states in $\mathcal{K}$.
These contributions are only functions of $k$ and $p$.

Then we see that the efficiency has the numerator and denominator:
\begin{align}
\label{memory-advantage-golden-mean}
\Delta_q C =& \frac{1}{Z}\left(C_\mu^{(\mathcal{K})}-C_q^{(\mathcal{K})}\right)
~\text{and}\\
\label{work-advantage-golden-mean}
\frac{\Delta_q W}{\kB T\ln 2} =& \frac{1}{Z}\left(C_\mu^{(\mathcal{K})}-C_q^{(\mathcal{K})}\right)
+\frac{(k-1)(1-p)-1}{Z} \Hq{S'|X=1}
  ~.
\end{align}
In this final form, we see that $Z$ cancels in the ratio $e_q = \kB T\ln 2
\Delta_q C/\Delta_q W$, and all that remains are functions that depend only on
$k$ and $p$.

\subsection{Nemo Generator}

For the Nemo Generator, we know that not all phases $\{\phi_{xs}\}$ give
equivalent implementations. To analyze the situation in more detail, we make use
of the gauge invariants.

The gauge invariants of the Nemo implementations are:
\begin{align}
\label{gauge-invar}
        \begin{split}
        \Phi_0 &= \phi_{0A} ~,\\
        \Phi_1 &= \phi_{1C}-\phi_{0C} ~,~\text{and}\\
        \Phi_2 &= \phi_{1A}+\phi_{1B}+\phi_{1C}
  ~.
        \end{split}
\end{align}
We work to express the overlap matrix in terms of these invariants.

Recall that the formula defining the overlaps. For the Nemo process, this gives
the system of equations:
\begin{align*}
\Omega_{AB} & =\sqrt{1-p} e^{i(\phi_{1C}-\phi_{1A})} ~,\\
\Omega_{BC} & =\frac{1}{\sqrt{2}} e^{i(\phi_{1C}-\phi_{1B})} ~,~\text{and}\\
\Omega_{CA} & =\sqrt{\frac{p}{2}}e^{i(\phi_{0A}-\phi_{0C})}+\sqrt{\frac{1-p}{2}}e^{i(\phi_{1A}-\phi_{1C})}\Omega_{AB}
  ~,
\end{align*}
which has the solution:
\begin{align*}
\Omega_{AB} & =\frac{\sqrt{p(1-p)}}{1+p} e^{i(\phi_{1C}-\phi_{1A}+\phi_{0A}-\phi_{0C})} ~,\\
\Omega_{BC} & =\frac{\sqrt{p}}{1+p} e^{i(\phi_{1C}-\phi_{1A}+\phi_{0A}-\phi_{0C})} ~,~\text{and}\\
\Omega_{CA} & =\frac{\sqrt{2p}}{1+p}e^{i(\phi_{0A}-\phi_{0C})}
		~.
\end{align*}
Now, we gauge fix $\phi_{1A}$ and $\phi_{1B}$ so that $\Omega_{AB}$ and
$\Omega_{BC}$ are phaseless. The result is:
\begin{align}
        \begin{split}
        \Omega_{AB} &=\frac{\sqrt{p(1-p)}}{1+p}  ~,\\
        \Omega_{BC} &=\frac{\sqrt{p}}{1+p}  ~,~\text{and}\\
        \Omega_{CA} &=\frac{\sqrt{2p}}{1+p}e^{i(2\Phi_0+2\Phi_1-\Phi_2)}
		~.
        \end{split}
\end{align}
We see that the overlap matrix then only depends on the gauge invariants in the
single phase $\Phi = 2\Phi_0+2\Phi_1-\Phi_2$. This generalizes a result from
Ref. \cite{Maho16a} to all input phases $\{\phi_{xs}\}$.

\end{document}